\documentclass[times,final,authoryear]{aastex63}

\shorttitle{Project Lyra: Jupiter Encounter}
\shortauthors{Hibberd}

\usepackage{graphicx}
\usepackage{rotating}
\usepackage{epstopdf}
\usepackage{epsf}
\usepackage{longtable}
\usepackage{savesym}
\savesymbol{tablenum}
\usepackage{siunitx}
\restoresymbol{SIX}{tablenum}
\usepackage{amssymb}
\usepackage{gensymb}
\usepackage{latexsym}
\usepackage{wrapfig}

\usepackage{url}
\usepackage{xcolor}
\definecolor{newcolor}{rgb}{.8,.349,.1}
\usepackage{bm}
\expandafter\ifx\csname package@font\endcsname\relax\else
 \expandafter\expandafter
 \expandafter\usepackage
 \expandafter\expandafter
 \expandafter{\csname package@font\endcsname}

\fi
\def\bq{\begin{equation}}
\def\eq{\end{equation}}
\def\bqy{\begin{eqnarray}}
\def\eqy{\end{eqnarray}}

\def\p{\partial}

\def\p{\partial}

\def\R{\mathbb{R}}

\graphicspath{{./}{figures/}}
\begin{document}
\title{\large{Project Lyra: Catching 1I/'Oumuamua Realistically  with a Jupiter Encounter and Imminent Propulsion Options }}

\author[0000-0003-1116-576X]{Adam Hibberd}
\affiliation{Initiative for Interstellar Studies (i4is),\\ 27/29 South Lambeth Road,\\ London SW8 1SZ, United Kingdom}



\begin{abstract}

Any missions to catch 1I/'Oumuamua have the daunting challenge of generating higher hyperbolic excess speeds than the interstellar object's own, i.e. $>$ 26.3 \si{km.s^{-1}} with respect to the Sun. To accomplish this task using chemical propulsion, previous papers have investigated a Solar Oberth manoeuvre and alternatively a Jupiter Oberth,  these options requiring a thrust from the chemical rocket at perihelion/perijove respectively, points at which the available velocity increment ($\Delta V$) results in maximum augmentation of the kinetic energy of the spacecraft. In this paper we unravel the specifics of a mission requiring a JOM or optionally a passive Jupiter encounter, i.e. with no thrust, the latter having so far not been addressed by Project Lyra. Whereas the previous papers were feasibility studies, this paper delves into what would be achievable with a Jupiter encounter (without any preceding gravitational assists from the inner planets), using currently available off-the-shelf solid and liquid rocket stages and assuming the NASA Space Launch System Block 2 can be deployed. Optimal Launch dates are found to lie in the years 2030, 2031 and 2032 with resulting mission durations of around 30-40 years, depending on the particular number and combination of stages exploited. Payload masses to 'Oumuamua of 860kg can readily be accomplished with a duration of 43 or so years, assuming a Centaur D and STAR 48B combination, the latter booster being ignited at perijove. On the other hand if two Centaur Ds are utilised instead of just one, retaining the STAR 48BV for the JOM, 35 years are evinced for the same mass. Furthermore, for lower payloads of $\sim{100}$kg, the flight duration can be cut accordingly to 31 years, with the additional benefit of requiring no burn at Jupiter.

\end{abstract}

\keywords{Project Lyra}


\section{Introduction}
\label{sec1}
In 2017, the first interstellar object to be discovered passing through our solar system, 1I/'Oumuamua, provoked consternation and excitement which reverberated around the scientific community, stimulating a great deal of research and conjecture concerning its nature and origin \citep{bannister2019natural}.\\

Examination of its light curve indicated two additional features which mark it out as quite unusual, firstly its tumbling rotational state and secondly the high ratio in its two extreme dimensions of around 5:1 or higher \citep{Siraj2019}, indicating a long cigar shape or more likely a flat pancake shape, the latter would not require any special orientation of 'Oumuamua with respect to the Earth \citep{Mashchenko_2019}. \\

Analysis of its trajectory as it sped away from the Earth revealed the presence of an anomalous force which could not be attributed to the gravitational attraction of the bodies of the solar system, and tailed off as 'Oumuamua receded from the sun according to the inverse or inverse square of its sun-distance \citep{Micheli2018}. If 'Oumuamua were like a typical solar system comet, such a force would be a sign of out-gassing, i.e. ejection of material due to sublimation of volatiles, yet it displayed absolutely no coma or tail. An alternative explanation might be solar radiation pressure and there have been various theories postulating that this might indeed be the cause, for example an alien solar sail or cometary dust bunny \citep{Bialy_2018,https://doi.org/10.48550/arxiv.2008.10083}.\\

Many theories exist attempting to resolve the mystery of 'Oumuamua, nearly all of which resort to hitherto unobserved phenomena to explain its unique properties - for example refer to \cite{Seligman_2020,Jackson2021,Desch2021,Raymond2018}.\\
 
Let us end this speculation - at the very least a simple close-up picture of the object would resolve all of this wild debate. There are obstacles in that it is out of range of both Earth-based and space-based telescopes (including the James Webb Space Telescope), and it is doubtful whether any near-future telescope could image it as its distance increases and it gets progressively fainter. However a spacecraft equipped with suitable spectrometers, with reference to appropriate empirical data sets, could potentially ascertain the composition of 'Oumuamua and also distinguish between a natural or artificial origin for the object \citep{https://doi.org/10.48550/arxiv.2211.02120}. Note that the \emph{Comet Interceptor} mission 
 \citep{S_nchez_2021} would be unable to intercept 'Oumuamua, due to its large and rapidly increasing radial distance from the sun and its accompanying displacement from the ecliptic plane. The architecture proposed in \cite{MOORE2021105137} is intended as a quick response mission and so clearly inappropriate for 'Oumuamua as the target.\\

\begin{wrapfigure}{l}{0.4\textwidth}
    \begin{center}
        \includegraphics[width=0.38\textwidth]{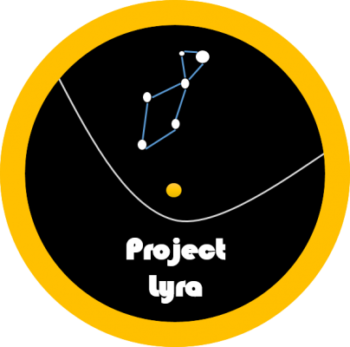}
    \end{center}
  \caption{\textit{Project Lyra} Logo, Initiative for Interstellar Studies (i4is)}
  \label{fig:PLyra}
 \end{wrapfigure}
 
Many scientists would wait for the Vera C. Rubin telescope to start operation before committing to a mission, as predictions indicate there will be further such objects arriving in our solar system and detectable by the telescope. An estimate of around $\sim{15}$ over a period of 10 years has been derived \citep{Hoover_2022}, whereas \cite{Marshall_2021} expect $\sim{7}$ to arrive within the orbit of Earth annually. Note that both of these papers base their number density of 'Oumuamua type objects in interstellar space on that determined by \cite{Do_2018}, which is an \emph{approximate upper bound} and there could be many fewer than these predictions would indicate.
Furthermore the scientific return achievable by a spacecraft mission, \textit{Project Lyra} (refer Figure \ref{fig:PLyra}), makes this an alternative and compelling option, especially in light of the inevitable passage of the spacecraft beyond the heliopause and into the Interstellar Medium (ISM). The latter is precisely the aim of the Interstellar Probe project propounded by the JHU APL \citep{9438249,MWG22}, and would of itself result in pivotal advances in our understanding of the composition of the ISM, which in turn would stand humanity in good stead for what should be our long term goal of exploration beyond the confines of our own home system. At this juncture, a little history of the Project Lyra research is in order.\\

In addition to explicating all the possible propulsion options which could be exploited by a spacecraft bound for 'Oumuamua (laser sails, solar sails, etc), the first Project Lyra paper \citep{HEIN2019552} also proposed a trajectory which utilised current or near-term chemical propulsion. This trajectory was solved by a software tool originally developed for preliminary interplanetary trajectory design, to wit \textit{Optimum Interplanetary Trajectory Software (OITS \cite{Hibberd_JBIS2022})} and adopted a \textit{Solar Oberth Manoeuvre (SOM)}, that is a slingshot of the sun, and a rocket burn at an extremely low perihelion - 3 solar radii - to take maximum advantage of the Oberth effect. However the research carried out therein concerning chemical propulsion options has now been rendered obsolete simply because the proposed launch date, in 2021, has now clearly elapsed.\\

A subsequent paper, \cite{Hibberd_2020}, revealed two important insights, firstly there were further launch windows beyond 2021 on a 12-yearly cycle (the nearest in the year 2033), and secondly that perihelia for the SOM could be increased without a pronounced and undesired reduction in the effectiveness of the SOM, thus relaxing the heat shield specification, and decreasing its required mass. The paper touched briefly upon the possibility of a simple powered Jupiter gravitational assist (or a \textit{Jupiter Oberth Manoeuvre, JOM}), and found a possible launch date in 2031. A further Project Lyra study adopted Nuclear Thermal Rockets as the propulsion method of choice \citep{HIBBERD2021594}, however it is the possibility of a mission undertaking a JOM alone, and using chemical, which this paper shall further investigate, as it represents a simple, clear and relatively direct alternative to the SOM (and also the VEEGA mission without a SOM - \cite{Hibberd_2022}), treated previously. In addition to a JOM, a passive Jupiter encounter is explored, an option so far not investigated in any previous Project Lyra papers.\\   
\section{Method}
\label{sec2}
For the theory behind OITS, please refer to \cite{Hibberd_JBIS2022}. For an explanation of the NLP (Non-Linear Programming) solvers used, i.e. NOMAD and MIDACO, refer to \cite{LeDigabel} and \cite{Schlueter_et_al_2009,Schlueter_Gerdts_2010,Schlueter_et_al_2009} respectively. This paper departs from previous Project Lyra studies because whereas in the past a top-down procedure was followed, i.e. trajectories were solved which minimised the total $\Delta V$ budget and thence a suitable combination of launcher and rocket boosters was derived using this available $\Delta V$; in the current paper an alternative, more logical bottom-up process is exploited where a combination of launcher and rocket boosters are selected, the consequent total $\Delta V$ is determined and a minimum \emph{flight time} trajectory is solved using OITS.\\

The reason for this new tack is as follows. To ensure the necessary extremely high reliability, chemical propellant boosters (liquid and solid), require significant investment to cover research, testing and developmental costs, but in comparison a relatively low outlay is needed in their manufacture and production. Thus for example commercially available solid boosters would be far more preferable than alternative bespoke ones developed from scratch. Because the performance of any booster is known and understood, it makes sense to execute a bottom-up approach to any research which adopts them.\\

A list of all the solid and liquid propellant stages utilised for the study is provided in Table \ref{RS_Info} and notice that the last two employ the $LH_{2}$/$LOX$ combination of liquid cryogens. For this reason they must actually be deployed shortly after injection by the launcher to minimise loss of $LH_{2}$ due to boil-off and leakage, and further without effective and weighty cryo-coolers which anyhow are currently unavailable, they could not be used for the JOM. Note also that the selection listed in Table \ref{RS_Info} is identical to that assumed for the Interstellar Probe research elaborated in \cite{MWG22}\\

\begin{table*}[]
\centering
\begin{tabular}{|c|c|c|c|c|c|}
\hline
\textbf{Rocket Stage} & \textbf{Solid/Liquid} & \textbf{Length (\si{m})} &\textbf{Total Mass (\si{kg})} & \textbf{Dry Mass (\si{kg})} & \textbf{Exhaust Velocity (\si{km.s^{-1}})} \\\hline
STAR 48B    & S & 2.03 & 2137  & 224 (+100)  & 2.8028 \\
ORION 50XL  & S & 3.07 & 4276(-30)  & 527 (+160)  & 2.8647 \\
CASTOR 30B  & S & 3.5  & 14521 (+550) & 1550 (+550) & 2.9649 \\
CASTOR 30XL & S & 6.0  & 27256 (+850) & 2242 (+850) & 2.8866 \\
CENTAUR III & L & 12.68 & 20830 & 3147 (+900) & 4.0090 \\
CENTAUR D   & L & 9.6  & 16458 (+200) & 2831 (+200) & 4.3512 \\ \hline
\end{tabular}
\caption{Rocket Stages Considered for Study, numbers in brackets represent differences from spec. needed to align performance with that given in Interstellar Probe Report \citep{MWG22}}
\label{RS_Info}
\end{table*}

\begin{figure}
  \centering
  \includegraphics[scale=0.6]{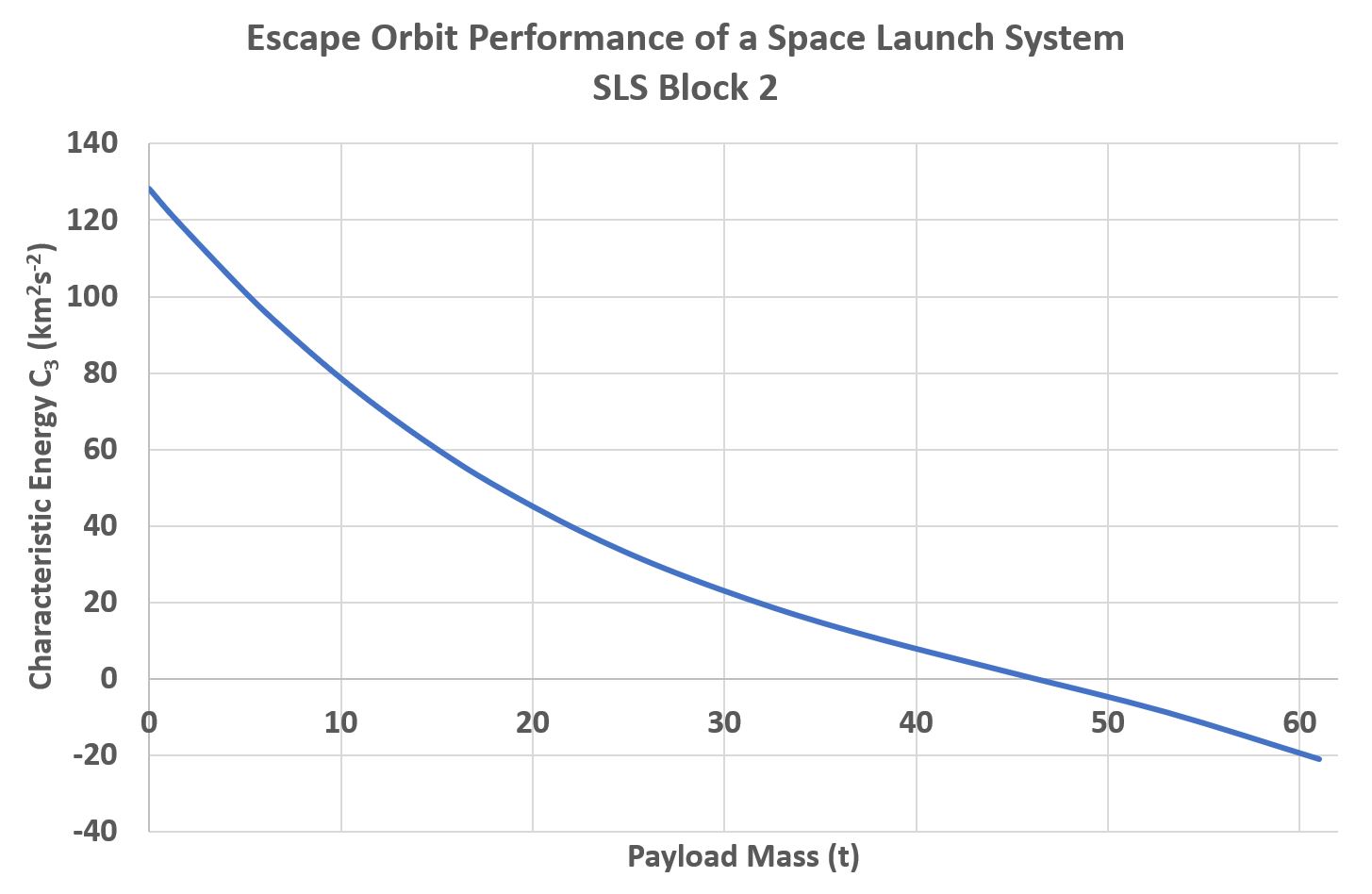}
  \caption{SLS Block 2 performance to Escape Orbit (derived from \cite{Stough2021NASAs,MWG22})}
  \label{fig:SLS}
 \end{figure}

As far as a launcher is concerned, to leverage the kinds of masses to Jupiter which would allow scope for the payload to include one or more rocket boosters given in Table \ref{RS_Info} so that a JOM can be executed, a super-heavy launch vehicle is necessary. There are two options which stand out over others, and they are the NASA Space Launch System (SLS) Block 2 and the SpaceX Starship. As there is copious data and specifics available online for the SLS Block 2 and relatively little on the Starship, and furthermore the SLS has achieved a successful maiden flight to orbit in 2022, for the analysis which follows the former of these two is chosen. The pertinent performance characteristics for the SLS Block 2 are encapsulated in Figure \ref{fig:SLS} where the \emph{Characteristic Energy $C_{3}$} is provided against mass of payload. This Figure \ref{fig:SLS} is derived from \cite{Stough2021NASAs,MWG22}.\\

This performance can be enhanced by incorporating a further stage or two after launcher injection and the $C_{3}$ value can be correspondingly amplified for any payload mass by such an arrangement.\\

 To calculate optimal time trajectories, OITS requires (a) constraint on $V_{\infty}$ at Earth and (b) constraint on $\Delta V_{perijove}$ at Jupiter (for the JOM). The former of these represents the \emph{hyperbolic excess speed} (for OITS this is also the speed at which the spacecraft leaves the Earth's gravitational sphere of influence) which in turn needs to be delivered by the combination of SLS Block 2 and rocket booster stages ignited after stage 2 burn-out. The following algorithm computes these two values (a) and (b):
 
 \begin{enumerate}
    \item adopt an eventual payload mass $M_{PAY}$ to be sent to 'Oumuamua after the Jupiter encounter
    \item select from Table \ref{RS_Info} a combination of stages to be fired at Earth $S_{i}$, $i=1,...,N_{E}$ and also a combination to be fired at the JOM $S_{i}$, $i=N_{E}+1,...,N$
    \item compute the sum $M_{TOT} = M_{PAY}+ \Sigma{_{i=1,N} M_{i}}$
    \item find the achievable SLS Block 2 $C_{3}$ capability from Figure \ref{fig:SLS} corresponding to $M_{TOT}$
    \item determine the speed of the SLS Block 2 at injection, $V_{P}$, equivalent to this $C_{3}$ value, i.e. $V_{P}^2 = C_{3} + V_{ESC}^2$ where $V_{ESC}^2 = 121.19  \si{km^2.s^{-2}}$ 
    \item from Tsiolkovsky compute the $\Delta V_{E}$ which the additional stages fired at Earth $S_i$ $i=1,...,N_E$ can generate from the specs given in Table \ref{RS_Info} and add this to $V_P$, i.e. $V_P'=V_P + \Delta V_{E}$
    \item compute the total $C_3'$ generated by SLS Block 2 and all these stages $C_{3}' =V_P'^2 - V_{ESC}^2$
    \item thus the hyperbolic excess, i.e. (a) above, is $V_{\infty} = \sqrt{C_3'}$
    \item $\Delta V_{perijove}$, i.e. (b) is calculated from Tsiolkovsky for the remaining stages at Jupiter $S_{i}$, $i=N_{E}+1,...,N$ using payload mass $M_{PAY}$ 
   \end{enumerate}
   
\section{Results}
\subsection{Pork Chop Plots}

Before we embark on the procedure set forth in Section \ref{sec2}, it is instructive to examine the general feasibility of a mission to 'Oumuamua using a JOM, for a period spanning the years 2025-2040, with particular attention to the '30s when the SLS block 2 should be operative and furthermore there is sufficient time to prepare a mission.\\

\begin{figure*}
\centering
\gridline{\fig{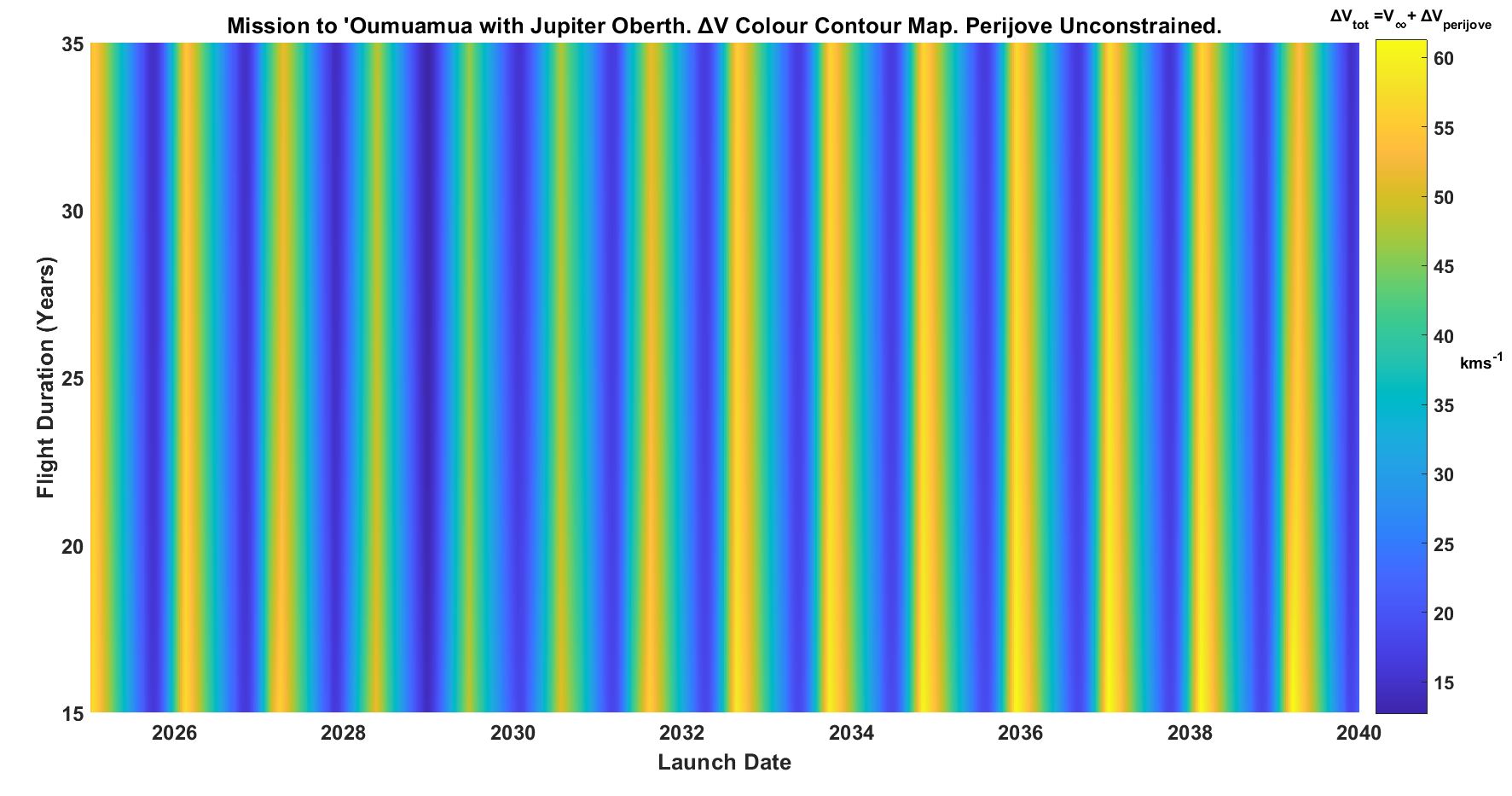}{1.0\textwidth}{(a) No perijove Constraint Imposed.}
          }
\gridline{\fig{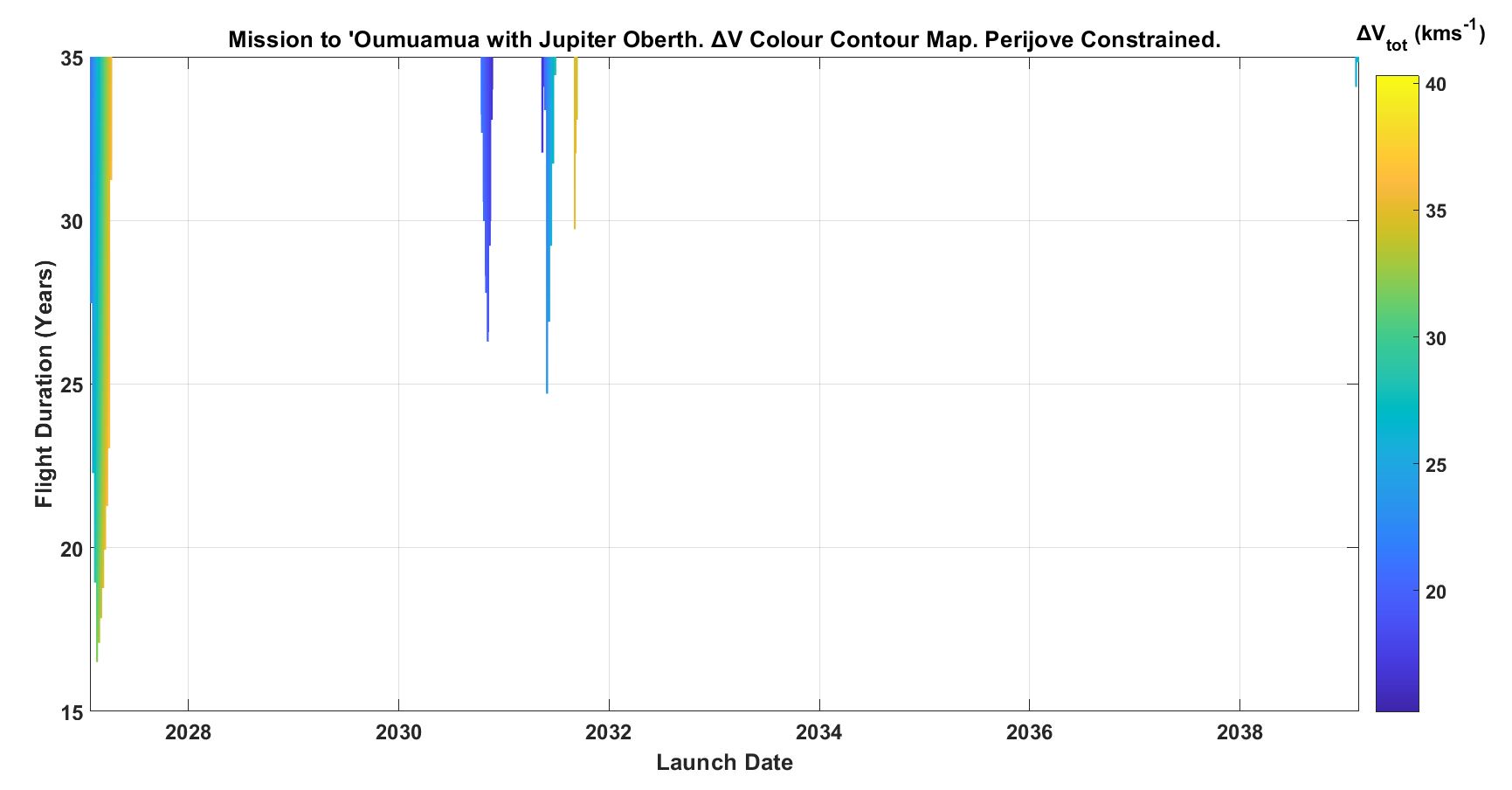}{1.0\textwidth}{(b)Landscape removed from (a) to allow positive perijove Altitudes only and viable perihelia for both legs of the journey (Earth-Jupiter \& Jupiter-'Oumuamua).}
          }
\caption{Pork Chop Colour Contour Map of Total $\Delta V$ as a Function of Launch Date and Total Flight Duration, Where the ratio of Flight Time from Earth to Jupiter and from Jupiter to 'Oumuamua is Optimal.
\label{fig:PorkChop}}
\end{figure*}

To this end refer to Figure \ref{fig:PorkChop} (a). This colour contour map provides colours indicating the magnitude of $\Delta V_{tot}$ = $V_{\infty}$ + $\Delta V_{perijove}$, where deep blue colours are low values, corresponding to greater feasibility and the brighter, yellower colours are associated with higher values, i.e. lower feasibility.\\

One can observe an unmistakable yearly pattern along the horizontal axis (launch date) when the Earth occupies yearly optimal points for reaching Jupiter in its orbit around the sun. Note that this plot is not the full story because although it neatly illustrates the celestial alignments necessary for optimal conditions to 'Oumuamua, nevertheless most of this landscape suffers from negative perijove altitudes, i.e. the perijove is unconstrained.\\

For a more representative analysis, refer to Figure \ref{fig:PorkChop} (b) which blanks out all regions of negative perijove altitude and also all perihelia which are infeasibly low. It is clear from this plot that before 2040 and after 2029, the only viable missions occur in three spikes around 2030, 2031 and 2032. In the following analysis therefore it makes sense to restrict the launch windows to around 2030, 2031 and 2032.\\

In the subsequent sections we shall endeavour a more detailed analysis of missions to 'Oumuamua using a Jupiter encounter, concentrating on the specific years 2030, 2031 and 2032.\\

\subsection{Two Example Mission Scenarios}{\label{Ex_Miss}}
In this section we concentrate on the specifics of a mission to 'Oumuamua using:
\begin{enumerate}
\item{Passive Jupiter GA \textit{without} thrust}
\item{A Powered Jupiter GA \textit{with} thrust (i.e. a JOM)}
\end{enumerate}
We re-iterate here that the NASA SLS Block 2 is chosen for the launch vehicle in question, and further there are two additional stages, a CENTAUR III followed by a STAR 48B.\\

\begin{figure}
\centering
  \includegraphics[scale=0.55]{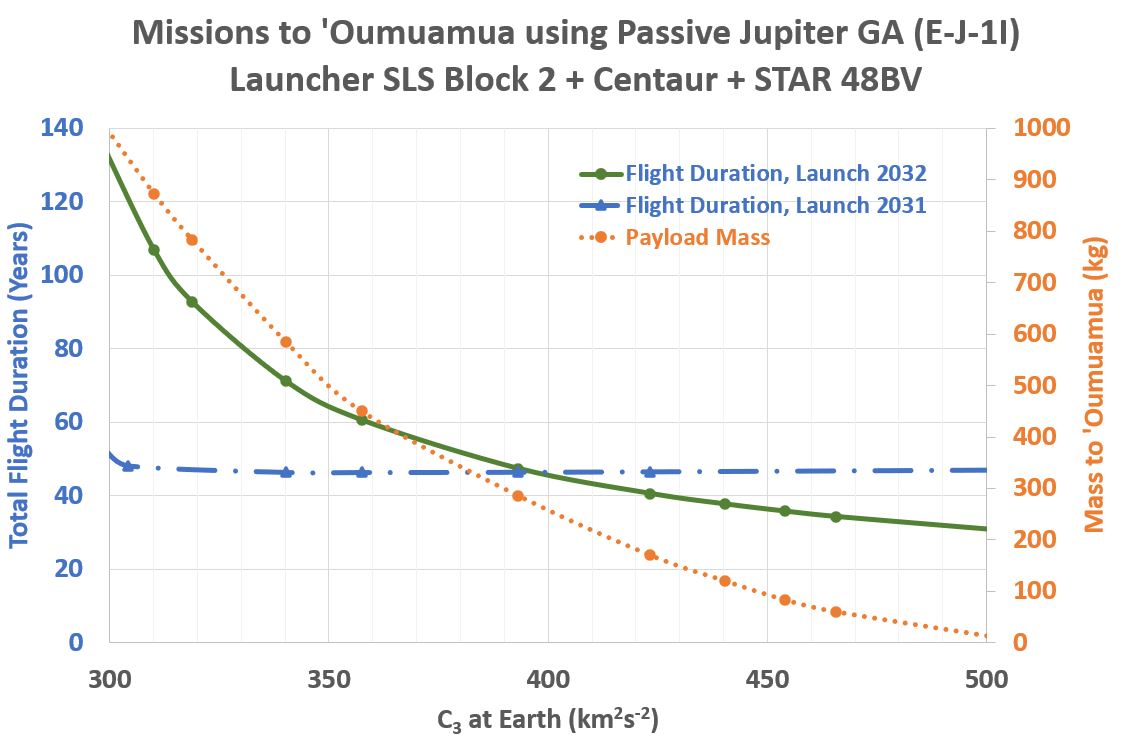}
  \caption{Passive Jupiter Encounter Results for CENTAUR III and STAR48B}
  \label{fig:PJE}
 \end{figure}
 
\begin{figure}
\centering
  \includegraphics[scale=0.55]{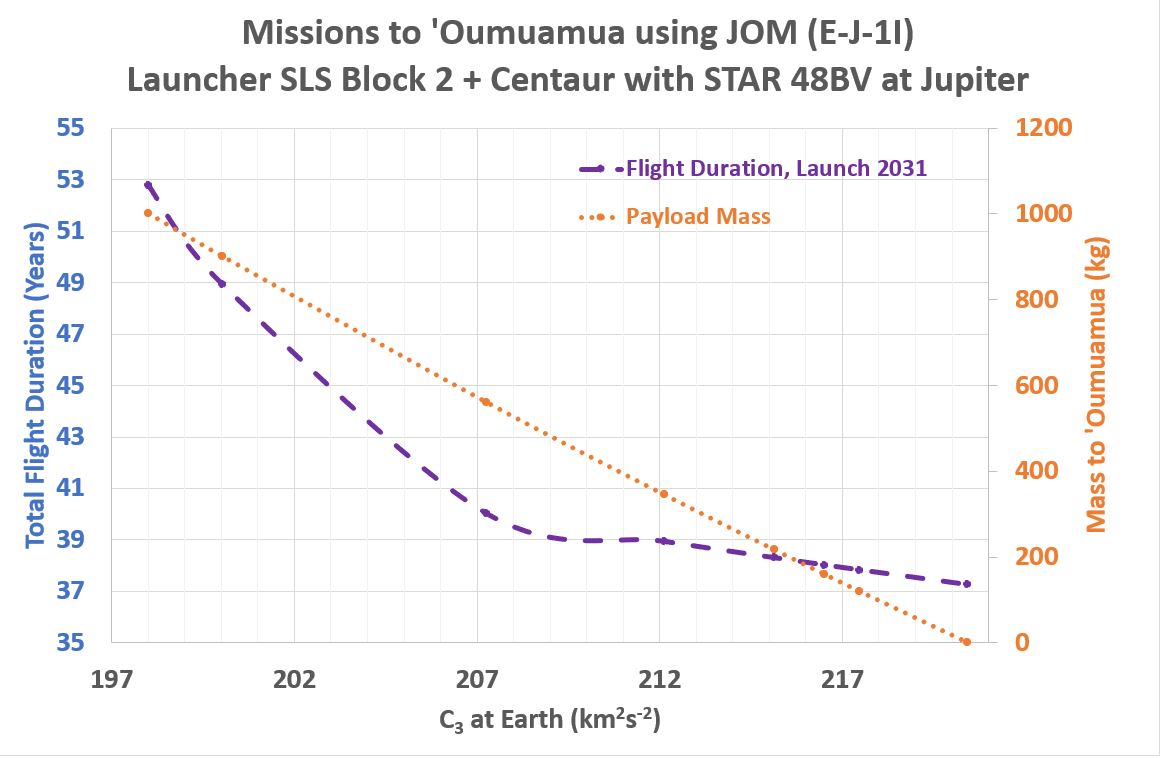}
  \caption{Powered Jupiter Encounter Results for CENTAUR III and STAR48B}
  \label{fig:JOM}
 \end{figure}

 \begin{figure}
\centering
  \includegraphics[scale=0.55]{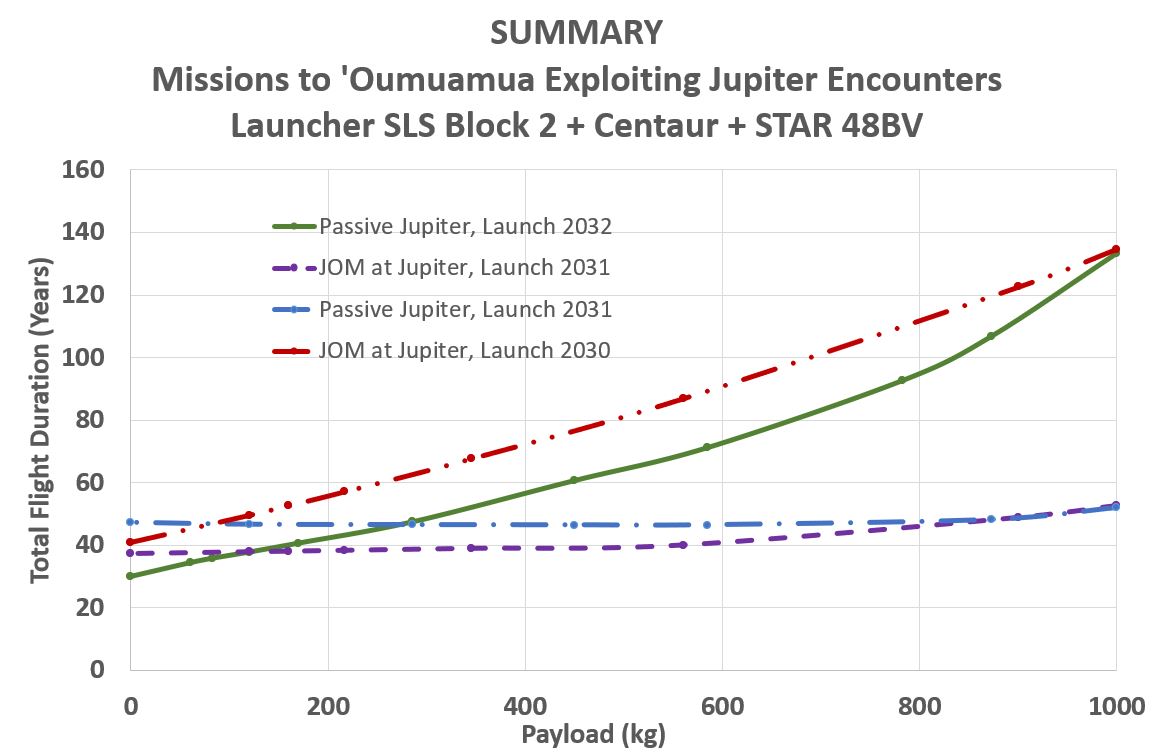}
  \caption{Summary of Jupiter Encounter Results for CENTAUR III and STAR48B}
  \label{fig:DrM}
 \end{figure}
 
In order to cater for scenarios (1) \& (2) above, we consider case (1) by firing both the CENTAUR III and immediately subsequently the STAR 48B at Earth injection (with no perijove burn), and case (2) by delaying the firing of the STAR 48B until Jupiter perijove. Indeed these two options are precisely those elaborated in the Interstellar Probe Report \citep{MWG22}.\\

Figure \ref{fig:PJE} provides the results of a passive Jupiter encounter for launches in the years 2031 and 2032. \emph{It was found that passive Jupiter encounters are infeasible for launches in 2030.} Observe that for payloads less than around 230 \si{kg} (i.e. $M_{PAY} < 230  \si{kg}$), a 2032 launch is clearly preferred, reaching a flight duration of 37 years for $M_{PAY}$ = 100 \si{kg}. Furthermore for a launch in 2031 and payload masses above 230 \si{kg}, the flight duration remains pretty well static at 46 or so years, whatever payload mass is chosen.\\

To compare this with a JOM, and a 2031 launch, refer Figure \ref{fig:JOM}. \emph{It was found that missions with a JOM launched in 2032 lead to flight durations on the order of $10^3$ $\si{years}$ and so can clearly be discounted.} For a JOM and launch in 2030, refer to Figure \ref{fig:DrM} which also includes all other viable combinations of launch years and trajectory profiles using the same two boosters mentioned above.\\

To summarise, it appears for $M_{PAY}$ $<$ 100 \si{kg} that a passive launch in 2032 is favoured, leading to flight durations of $\sim{35} \si{years}$, whereas above this mass, the JOM with a launch in 2031 is preferred, giving durations of $\sim{40} \si{years}$ fairly consistently up to $M_{PAY} \sim{900}\ \si{kg}$.\\

\subsection{Investigation of Different Payload Masses}

First we choose the payload mass adopted by the \textit{Interstellar Probe} Report of 860 \si{kg}. (Go to Section \ref{disc} for an overview of the similarities and differences of a mission to 'Oumuamua compared to the Interstellar Probe.) \\

For a summary of the results, refer to Table \ref{M860kg}. We find the optimal trajectory highlighted in this table is a JOM with launch in 2031 and an associated overall flight duration of 35.2 years. Two Centaur D stages are fired consecutively after Earth injection and then following this, the STAR 48B is fired at Jupiter perijove. For information, the velocity of the spacecraft relative to 'Oumuamua upon arrival is around 11.4 \si{km.s^{-1}}.\\

\begin{table*}
\centering
\begin{tabular}{|c|c|c|c|c|c|c|c|c|c|}
\hline
\textbf{Trajectory} & \textbf{JOM/} & \textbf{Launch} & \textbf{Stage 1} & \textbf{Stage 2} & \textbf{Stage 3} & \textbf{Duration}  & \textbf{Declin.} & \textbf{Perijove} & \textbf{Sun-Dist.}\\
\textbf{Option} & \textbf{Passive} & \textbf{Year} & \textbf{} & \textbf{} & \textbf{} & \textbf{(Years)} & \textbf{(\degree)} & \textbf{alt. (km)} & \textbf{(au)} \\ \hline
EJJ & JOM & 2031 & CENTAUR D & STAR 48B & STAR 48B & 52.6 & -22.85 & 44154 & 373\\
EJJ & JOM & 2030 & CASTOR 30XL & ORION 50 XL & ORION 50 XL & 50.3 & -7.31 & 200 & 354\\
EEE & Passive & 2032 & CENTAUR D & CENTAUR D & STAR 48B & 63.2 & -22.69 & 125481 & 436 \\
EEE & Passive  & 2031 & CENTAUR D & CENTAUR D & STAR 48B & 46.4 & -22.15 & 200 & 338\\
EEJ & JOM & 2030 & CENTAUR D & CENTAUR D & CASTOR 30B & 47.3 & -22.92 & 200 & 338\\ \hline
\textbf{EEJ} & \textbf{JOM} & \textbf{2031} & \textbf{CENTAUR D} & \textbf{CENTAUR D} & \textbf{STAR 48B} & \textbf{35.2} & \textbf{-22.92} & \textbf{3435} & \textbf{277} \\ \hline
EE & Passive  & 2032 & CENTAUR D & STAR 48B & N/A & 89 & -23.08 & 157276 & 581\\
EE & Passive  & 2031 & CENTAUR D & STAR 48B & N/A & 46.6 & -22.65 & 200 & 340\\
EJ & JOM & 2031 & CENTAUR D & STAR 48B & N/A & 43.4 & -22.89 & 18988 & 322\\
EJ & JOM & 2030 & CENTAUR D & CASTOR 30B & N/A & 68.3 & -7.12 & 24086 & 455\\ \hline
\end{tabular}
\caption{Optimal stage configurations and flight durations for a payload mass of \textbf{860 \si{kg}}.}
\label{M860kg}
\end{table*}
We now constrain the number of stages to 2. We find that the JOM with a launch in 2031 turns out to be a marginal improvement on a passive encounter (launched in the same year), at 43.4 years for the former compared to 46.6 years for the latter.\\

\begin{figure}
  \includegraphics[angle=-90,scale=0.82]{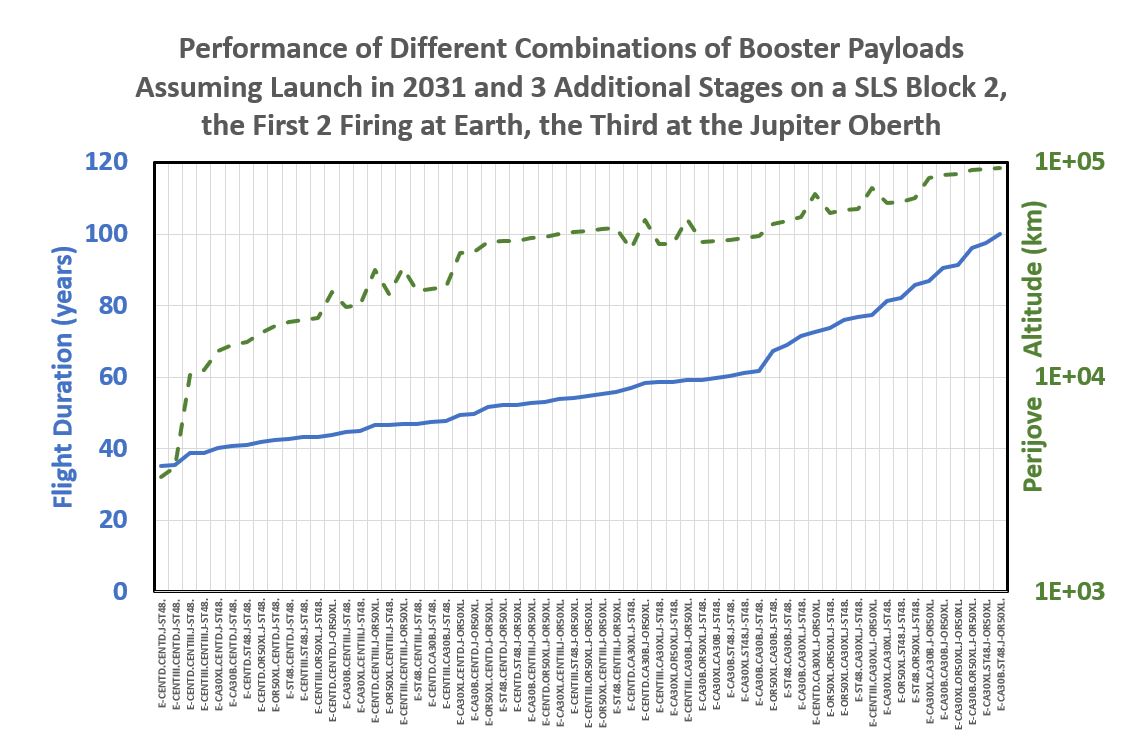}
  \caption{Summary of all studied combinations of boosters assuming the EEJ + 2031 combination as given in Table \ref{M860kg}}
  \centering
  \label{fig:EEJSum}
\end{figure}

A summary of all possible permutations of the rocket stages (as is given in Table \ref{RS_Info}) is provided in Figure \ref{fig:EEJSum} which assumes a launch in 2031 and a mass 860 \si{kg} payload, and further discounts all those instances where the last stage is either a Centaur III or Centaur D (because without currently unavailable cryo-coolers they cannot be fired at the JOM due to $LH_2$ boil-off and leakage en route to Jupiter).\\

On the other hand if instead a payload mass comparable to the \emph{New Horizons} spacecraft to Pluto - i.e. 480 \si{kg} - is selected, and the same 3 stage scenario as is optimal for the 860 \si{kg} case is chosen, then it is found that their is a marginal improvement on flight duration at around 33.5 \si{years} compared to the aforementioned 35.2 \si{years}.\\

Finally for purposes of comparison if we assume a payload mass as low as 100 \si{kg}, with the same scenario highlighted in Table \ref{RS_Info}, i.e. a launch in 2031 with a JOM, then investigation reveals the flight duration drops, by only a small amount, to 33.4 \si{years}, compared to the previously mentioned 33.5 \si{years} with the 480 \si{kg} option. However, from previous work which was detailed in Section \ref{Ex_Miss}, it would appear that a 2032 launch with passive Jupiter evinces lower flight durations for payload masses of around 100 \si{kg}. Accordingly, when the analysis is conducted for this mass, a flight duration as low as 31.0 \si{years}, which is the lowest flight duration so far, is indicated.\\

To summarise the results of this section, down to a mass of around  $M_{PAY} \sim{100} \si{kg}$, there appears to be no obvious benefit in terms of flight duration to 'Oumuamua in choosing lower payload masses, like for example a New Horizons size spacecraft, over one with a larger size, as adopted for the Interstellar Probe, assuming in both cases the optimal scenario is chosen of a launch in 2031 with two Centaur D fired at Earth and a further thrust is applied at Jupiter by a STAR 48BV.\\

However for a mass lower than $M_{PAY} \sim{100} \si{kg}$, we find a passive Jupiter and launch in 2032 is preferable, enabling a mission of 31 \si{years} in duration.

\section{Discussion}\label{disc}
Here we shall not formulate specific mission architectures based on the discoveries already elucidated. For a detailed description of a treatment of architectures, refer to \cite{Bannova_22}.\\

First of all, there are two trajectory options which stand out as possibilities for a mission:

\begin{enumerate}
    \item{firstly a launch in the year 2031, either a JOM or passive Jupiter encounter,}
    \item{a passive Jupiter in the year 2032.}
\end{enumerate}    
    
 The preferred option of these is dependent on the payload mass budget available for the mission.\\
 
 If for example a mass equivalent to the Interstellar Probe (860kg) is deemed necessary then the first option has very much the superior performance in terms of minimising flight duration, the JOM allowing a flight duration of 35.2 years. Furthermore when this mass is reduced, it seems this advantage is maintained down to a payload mass of $\sim{130}$ kg, without any appreciable reduction in flight duration. Below this latter mass the second of the two options becomes preferable, allowing flight durations as low as 31 years.\\

 In light of alternative future propulsion systems such as laser sails \citep{Turyshev2020}, which would necessitate technologies for the Breakthrough Starshot Initiative \citep{PARKIN22} to be in place, the infrastructure for such capabilities may be available at a basic level some time in the mid '30s. If we take fiducial parameters from \cite{Turyshev2020} entailing a spacecraft mass of 100 kg and a speed generated of $\sim{300 \si{km.s^{-1}}}$, a mission to 'Oumuamua using this architecture would take on the order of $\sim$ a few years \citep{HIBHEIN}. In principle then, if we invoke the \emph{incentive trap} \citep{Kennedy_2010}, this would compel us to wait until such a mission can be seriously realised, assuming the rate of technological growth in this area conforms with current expectations.\\
 
 However note that the Interstellar Probe initiative, with a mass of 860kg and equipped with all the necessary instrumentation to study the heliosphere and interstellar medium \citep{MWG22}, also assumes chemical propulsion and a launch in the early '30s on an SLS Block 2. In addition it concludes that a passive Jupiter or JOM would be the baseline mission scenario for the project, the former taking precedence. Thus as well as satisfying our curiosity around 'Oumuamua, such a mission would enable a pragmatic resolution to our questions surrounding our immediate interstellar neighbourhood.\\
 
 It should be noted also that a mission to 'Oumuamua would have significant differences from the Interstellar Probe mission. The main reason is that Project Lyra has a specific target to reach and so some form of terminal guidance would be required. Also due to uncertainties in 'Oumuamua's precise escape asymptote, a mission would have to travel along a \emph{best estimate}, which might diverge quite significantly from the true direction. Thus some form of target acquisition would need to be implemented. To this end a New Horizons type LORRI telescope would suffice \citep{Hein_2022}. Also appreciable trajectory adjustment would need to be present via an on-board propulsion system.       

\section{Conclusion}

The Jupiter encounter option has an appeal in that it is very much tried and tested technology, requiring no heat shield as would be necessary for the alternative of a Solar Oberth Manoeuvre (SOM). The flight durations are accordingly longer than the SOM, by typically 10 years or more, but the option has a higher mass capability than the SOM. Either a launch in 2031 is indicated or a launch in 2032 would be possible where a passive Jupiter encounter can be exploited. Flight Durations of 31 to 45 or so years can be realised.

\bibliography{Using_Jupiter,library_Adam_Hibberd,Hein_ISO_Modified_by_Adam_Hibberd}{}
\bibliographystyle{aasjournal}



\end{document}